\documentclass{aa}  

\usepackage{amsmath}
\usepackage{soul}
\usepackage{graphicx}
\usepackage{natbib}
\usepackage{bm}
\usepackage{url}
\usepackage{color}
\usepackage{txfonts}
\usepackage{caption}
\usepackage{subcaption}
\usepackage{appendix}
\usepackage{rotating}
\usepackage{longtable}
\usepackage{booktabs}
\usepackage{lscape}
\usepackage[table]{xcolor}
\begin{document}

    \title{Extended transition rates and lifetimes in \ion{Al}{I} and \ion{Al}{II} from systematic multiconfiguration calculations
    	}

    \author{A. Papoulia\inst{1,2}
          \and J. Ekman\inst{1}
          \and P. J\"{o}nsson\inst{1} }

   \institute{Materials Science and Applied Mathematics, Malm\"{o} University, SE-20506 Malm\"{o}, Sweden \\
             \email{asimina.papoulia@mau.se} 
   \and
         Division of Mathematical Physics, Lund University, Post
         Office Box 118, SE-22100 Lund, Sweden\\
             }
%\titlerunning 
%\authorrunning
   \date{Received -, 2018; }

\abstract{}{The objective of this work is to provide a substantial amount of updated atomic data for the systems of neutral and singly ionized aluminium, including transition data in the infrared region. This is particularly important since the new generation of telescopes are designed for this region.}{Multiconfiguration Dirac-Hartree-Fock (MCDHF) and relativistic configuration interaction (RCI) calculations were performed for 28 and 78 states in neutral and singly ionized aluminium, respectively. In \ion{Al}{I}, the configurations of interest are $3s^2nl$ for $n=3,4,5$ with $l=0$ to $4$, as well as $3s3p^2$ and $3s^26l$ for $l=0,1,2$. In \ion{Al}{II}, the studied configurations are, besides the ground configuration $3s^2$, $3snl$ with $n=3$ to $6$ and $l=0$ to $5$, $3p^2$, $3s7s$, $3s7p$ and $3p3d$. Valence and core-valence electron correlation effects are systematically accounted for through large configuration state function (CSF) expansions.}{Calculated excitation energies are found to be in excellent agreement with experimental data from the NIST database. Lifetimes and transition data for radiative electric dipole (E1) transitions are given and compared with results from previous calculations and available measurements, for both \ion{Al}{I} and \ion{Al}{II}. The computed lifetimes of \ion{Al}{I} are in very good agreement with the measured lifetimes in high-precision laser spectroscopy experiments.  There is a significant improvement in accuracy, in particular for the more complex system of neutral \ion{Al}{I}.}{} 
% 5 {} token are mandatory

%{Multiconfiguration Dirac-Hartree-Fock (MCDHF) and relativistic configuration interaction (RCI) calculations were performed for 28 and 78 states in neutral and singly ionized aluminium, respectively. In \ion{Al}{I}, the configurations of interest are $3s^2nl$ for $n=3,4,5$ with $l=0$ to $4$, as well as $3s3p^2$ and $3s^26l$ for $l=0,1,2$. In \ion{Al}{II}, the studied configurations are, besides the ground configuration $3s^2$, $3snl$ with $n=3$ to $6$ and $l=0$ to $5$, $3p^2$, $3s7s$, $3s7p$ and $3p3d$. Valence and core-valence electron correlation effects are systematically accounted for through large configuration state function (CSF) expansions. Calculated excitation energies are found to be in excellent agreement with experimental data from the NIST database. Lifetimes and transition data for radiative electric dipole (E1) transitions are given and compared with results from previous calculations and available measurements, for both \ion{Al}{I} and \ion{Al}{II}. The computed lifetimes of \ion{Al}{I} are in very good agreement with the measured lifetimes in high-precision laser spectroscopy experiments. The present calculations provide a substantial amount of updated atomic data, including transition data in the infrared region. This is particularly important since the new generation of telescopes are designed for this region. There is a significant improvement in accuracy, in particular for the more complex system of neutral \ion{Al}{I}. The complete tables of transition data are available online.}  


\titlerunning{Extended transition rates and lifetimes in \ion{Al}{I} and \ion{Al}{II}}

   \maketitle

%________________________________________________________________

\section{Introduction}

Aluminium is an important element in astrophysics. In newly born stars the galactic [\element{Al}/\element{H}] abundance ratio, as well as the [\element{Al}/\element{Mg}] ratio are found to be increased in comparison to early stars \citep{Clayton}. The aluminium abundance and its anti-correlation with the one of magnesium is the best tool to determine which generation a globular cluster star belongs to. The abundance variations of different elements and the relative numbers of first and second generation stars may be used to determine the nature of polluting stars, the timescale of the star formation episodes and the initial mass function of the stellar cluster \citep{Carretta}. The aluminium abundance is of importance for other types and groups of stars as well. A large number of spectral lines of neutral and singly ionized aluminium are observed in the solar spectrum and in many stellar spectra. Aluminium is one of the interesting elements for chemical analysis of the Milky Way and one example is the Gaia-ESO Survey\footnote{\url{http://casu.ast.cam.ac.uk/surveys-projects/ges}}, in which medium- and high-resolution spectra from more than $10^5$ stars are analyzed to provide public catalogs with astrophysical parameters. As part of this survey, \citet{GaiaESO} analyzed high-resolution UVES\footnote{\url{http://www.eso.org/sci/facilities/paranal/instruments/uves.html}} spectra of FGK-type stars and derived abundances for 24 elements, including aluminium. 

In addition, aluminium abundances have been determined in local disk and halo stars by \citet{Gehren}, \citet{Reddy}, \citet{Mishenina}, \citet{Adibekyan} and \citet{Bensby}. However, chemical evolution models still have problems reproducing the observed behavior of the aluminium abundance in relation to abundances of other elements. Such examples are the observed trends of the aluminium abundances in relation to metallicity [\element{Fe}/\element{H}], which are not well reproduced at the surfaces of stars, like giants and dwarfs \citep{GaiaSodAl}. On grounds of the above issues, \citet{GaiaSodAl} redetermined aluminium abundances within the Gaia-ESO Survey. Furthermore, strong deviations from local thermodynamic equilibrium (LTE) are found to significantly affect the inferred aluminium abundances in metal poor stars, which was highlighted in the work by \citet{Gehren2}. \citet{NordLind} presented a non-local thermodynamic equilibrium (NLTE) modeling of aluminium and provided abundance corrections for lines in the optical and near-infrared regions. 

Correct deduction of aluminium abundances and chemical evolution modeling is thus necessary to put together a complete picture of the stellar and Galactic evolution. Obtaining the spectroscopic reference data to achieve this goal is demanding. Significant amount of experimental research has been conducted to probe the spectra of \ion{Al}{II} and \ion{Al}{I} and facilitate the analysis of the astrophysical observations. Yet, some laboratory measurements still lack reliability and in many cases, especially when going to higher excitation energies, only theoretical values of transition properties exist. Accurate computed atomic data are therefore essential to make abundance analyses in the Sun and other stars possible.
 
For the singly ionized \ion{Al}{II}, there is a number of measurements of transition properties. The radiative lifetime of the $3s3p~^3P_1^o$ level was measured by \citet{Johnson} using an ion storage technique and the transition rate value for the inter-combination $3s3p~^3P_1^o~\rightarrow~3s^2~^1S_0$ transition was provided. \citet{Trabert} measured lifetimes in an ion storage ring and the result for the lifetime of the $3s3p~^3P_1^o$ level is in excellent agreement with the one measured by \citet{Johnson}. Using the beam-foil technique, \citet{Andersen} measured lifetimes for the $3snf~^3F$ series with $n=4-7$, although those measurements are associated with significant uncertainties. By using the same technique, the lifetime of the singlet $3s3p~^1P_1^o$ level was measured in four different experimental works \citep{Kernahan, Head, Berry, Smith}, which are in very good agreement. 

In the case of neutral \ion{Al}{I}, several measurements have also been performed. Following a sequence of earlier works \citep{JonssonLundberg, Jonsson}, \citet{Buurman} used laser spectroscopy to obtain experimental values for the oscillator strengths of the lowest part of the spectrum. A few years later, \citet{BuurDonsz} redetermined the lifetime of the $3s^24p~^2P$ level and separated the different fine-structure components. Using similar laser techniques, \citet{Davidson} measured the natural lifetimes of the $3s^2nd~^2D$ Rydberg series and obtained oscillator strengths for transitions to the ground state. In a more recent work, \citet{Vujnovic} used the hollow cathode discharge method to measure relative intensities of spectral lines of both neutral and singly ionized aluminium. Absolute transition probabilities were evaluated based on available results from previous studies, such as the ones mentioned above.

\ion{Al}{II} is a nominal two-electron system and the lower part of its spectrum is strongly influenced by the interaction between the $3s3d~^1D$ and $3p^2~^1D$ configuration states. Contrary to neutral \ion{Mg}{I} where no level is classified as $3p^2~^1D$, in \ion{Al}{II} the $3p^2$ configuration dominates the lowest $^1D$ term and yields a well-localized state below the $3s3d~^1D$ term. The interactions between the $3snd~^1D$ Rydberg series and the $3p^2~^1D$ perturber were investigated by \citet{TayalHibbert}. Going slightly further up, the spectrum of \ion{Al}{II} is governed by the strong mixing of the $3snf~^3F$ Rydberg series with the $3p3d~^3F$ term. Despite the widespread mixing, $3p3d~^3F$ is also localized, between the $3s6f~^3F$ and $3s7f~^3F$ states. The configuration interaction between doubly excited states, e.g. the $3p^2~^1D$ and $3p3d~^3F$ states, and singly excited $3snl~^{1,3}L$ states was thoroughly investigated by \citet{ChangWang}. However, the extreme mixing of the $3p3d~^3F$ term in the $3snf~^3F$ series and its effect on the computation of transition properties was first investigated by \citet{Weiss}. Although the work by \citet{ChangWang} was more of a qualitative nature, computed transition data were provided based on configuration interaction (CI) calculations. Using the B-spline configuration interaction (BSCI) method, \citet{ChangFang} also predicted transition properties, as well as lifetimes of \ion{Al}{II} excited states.  

Despite the large number of measured spectral lines in \ion{Al}{I}, the $3s3p^2~^2D$ state could not be experimentally identified and for a long time theoretical calculations had been trying to localize it and predict whether it lies above or below the first-ionization limit. \ion{Al}{I} is a system with three valence electrons and correlation effects are even stronger compared to the singly ionized \ion{Al}{II}. Especially strong is the two-electron interaction of $3s3d~^1D$ with $3p^2~^1D$, which becomes evident between the $3s^23d~^2D$ and $3s3p^2~^2D$ states. The $3s3p^2~^2D$ state is strongly coupled to the $3s^23d~^2D$ state, but it is also smeared out over the entire discrete part of the $3s^2nd~^2D$ series and contributes to a significant mixing of all those states \citep{Weiss}. Asking for the position of the $3s3p^2~^2D$ level is thus meaningless, since it does not correspond to any single spectral line \citep{Lin, Trefftz}. Due to this strong two-electron interaction, the line strength of one of the $^2D$ states involved in a transition appears to be enhanced, while the line strength of the other $^2D$ state is being suppressed. This makes the computation of transition properties in \ion{Al}{I} far from trivial \citep{Froese}. More theoretical studies on the system of neutral aluminium were conducted by \citet{Taylor} and \citet{Theodosiou}.
 
In view of the great astrophysical interest for accurate atomic data, close coupling (CC) calculations were carried out for the systems of \ion{Al}{II} and \ion{Al}{I}, by \citet{Butler} and \citet{Mendoza}, respectively, as part of the Opacity Project. These extended spectrum calculations produced transition data in the infrared region (IR), which were scarce until then. However, the neglected relativistic effects, as well as the insufficient amount of correlation included in the calculations constitute limiting factors to the accuracy of the results. Later on, \citet{Froese} performed multiconfiguration Hartree-Fock (MCHF) calculations and used the Breit-Pauli (BP) approximation to also capture relativistic effects for Mg- and Al-like sequences. Focusing more on correlation, relativistic effects were kept to lower order. Even so, in \ion{Al}{I}, correlation in the core and core-valence effects were not included due to limited computational resources. The latest compilation of \ion{Al}{II} and \ion{Al}{I} transition probabilities was made available by \citet{Kelleher}. \citet{WieseMartin} had earlier updated the first critical compilation of atomic data by \citet{Wiese1}.
  
Although for the past decades a considerable amount of research has been conducted for the systems of \ion{Al}{II} and \ion{Al}{I}, there is still a need for extended and accurate theoretical transition data. The present study is motivated by such a need. To obtain energy separations and transition data, the fully relativistic multiconfiguration Dirac-Hartree-Fock (MCDHF) scheme has been employed. Valence and core-valence electron correlation is included in the computations of both systems. Spectrum calculations have been performed to include the first 28 and 78 lowest states in neutral and singly ionized aluminium, respectively. Transition data corresponding to IR lines have also been produced. The excellent description of energy separations is an indication of highly accurate computed atomic properties, which can be used to improve the interpretation of abundances in stars.

\section{Theory}

\subsection{Multiconfiguration Dirac-Hartree-Fock}

The wave functions describing the states of the atom, referred to as atomic state functions (ASFs), are obtained by applying the multiconfiguration Dirac-Hartree-Fock approach (MCDHF) \citep{Gr, Froese1}. In the MCDHF method, the ASFs are approximate eigenfunctions of the Dirac-Coulomb Hamiltonian given by
\begin{equation}
{H}_{DC}= \sum_{i=1}^{N} [c \hspace{0.1cm} \bm\alpha_i\cdot \textbf{p}_i + (\beta_i - 1)c^2 + V_{nuc}(r_i)] + \sum_{i<j}^{N} \frac{1}{r_{ij}} ,\label{Eq1}
\end{equation}
where $V_{nuc}(r_i)$ is the potential from an extended nuclear charge distribution, $\bm{\alpha}$ and $\beta$ are the $4 \times 4$ Dirac matrices, $c$ the speed of light in atomic units and $\textbf{p} \equiv -\rm{i}\nabla$ the electron momentum operator. An ASF $\Psi(\gamma PJM_J)$ is given as an expansion over $N_{CSF}$ configuration state functions (CSFs), $\Phi(\gamma_i PJM_J)$, characterized by total angular momentum $J$ and parity $P$:
\begin{equation}
\Psi (\gamma PJM_J) = \sum_{i=1}^{N_{CSF}} c_i \Phi(\gamma_i PJM_J). \label{Eq2}
\end{equation}
The CSFs are anti-symmetrized many-electron functions built from products of one-electron Dirac orbitals and are eigenfunctions of the parity operator $P$, the total angular momentum operator $J^2$ and its projection on the $z$-axis $J_z$ \citep{Gr, Froese1}. In the expression above, $\gamma_i$ represents  the configuration, coupling and other quantum numbers necessary to uniquely describe the CSFs.

The radial parts of the Dirac orbitals together with the mixing coefficients $c_i$ are obtained in a self-consistent field (SCF) procedure. The set of SCF equations to be iteratively solved results from applying the variational principle on a weighted energy functional of all the studied states according to the extended optimal level (EOL) scheme \citep{Dyall}. The angular integrations needed for the construction of the energy functional are based on the second quantization method in the coupled tensorial form \citep{Gaigalas1, Gaigalas2}. 

The transverse photon (Breit) interaction, as well as the leading quantum electrodynamic (QED) corrections (vacuum polarization and self-energy) can be accounted for in subsequent relativistic configuration interaction (RCI) calculations \citep{McKenzie}. In the RCI calculations, the Dirac orbitals from the previous step are fixed and only the mixing coefficients of the CSFs are determined by diagonalizing the Hamiltonian matrix. All calculations were performed using the relativistic atomic structure package GRASP2K \citep{Jonsson1}.

In the MCDHF relativistic calculations, the wave functions are expansions over $jj$-coupled CSFs. To identify the computed states and adapt the labeling conventions followed by the experimentalists, the ASFs are transformed from $jj$-coupling to a basis of $LSJ$-coupled CSFs. In the GRASP2K code this is done using the methods developed by \citet{Gaigalas3, Gaigalas4,ATOMSJJ2LSJ}. 

\subsection{Transition parameters}

Besides excitation energies, also lifetimes $\tau$ and transition parameters, such as emission transition rates $A$ and weighted oscillator strengths $gf$, were computed. The transition parameters between two states $\gamma'P'J'$ and $\gamma PJ$ are expressed in terms of reduced matrix elements of the transition operator \textbf{T} \citep{Grant2}
\begin{equation}
\langle \Psi (\gamma PJ)||\textbf{T}||\Psi (\gamma' P'J') \rangle = \sum_{k,l} c_k c'_l \langle \Phi(\gamma_k PJ)||\textbf{T}||\Phi(\gamma'_l P'J') \rangle.
\end{equation}

For electric multipole transitions, there are two forms of the transition operator: the length, which in fully relativistic calculations is equivalent to the Babushkin gauge and the velocity form, equivalent to the Coulomb gauge. The transitions are governed by the outer part of the wave functions. The length form is more sensitive to this part of the wave functions and it is generally considered as the preferred form. Regardless, the agreement between the values of these two different forms can be used to indicate the accuracy of the wave functions \citep{Charlotte2, Ekman}. This is particularly useful when no experimental measurements are available. The transitions can be organized in groups, determined, for instance, by the magnitude of the transition rate value. A statistical analysis of the uncertainties of the transitions can then be performed. For each group of transitions, the average uncertainty to the length form of the computed transition rates is given by
\begin{equation}
\langle dT \rangle = \frac{1}{N} \sum_{i=1}^{N} \frac{|A^i_l - A^i_{\upsilon}|}{\textnormal{max}(A^i_l,A^i_{\upsilon})}, 
\end{equation}
where $A_l$ and $A_{\upsilon}$ are respectively the transition rates in length and velocity form for a transition $i$ and $N$ is the number of the transitions belonging to a group. In this work, we only computed transition parameters for the electric dipole (E1) transitions. The electric quadrupole (E2) and magnetic multipole (Mk) transitions are much weaker and therefore less likely to be observed.

\section{Calculations}

\subsection{Al I}

In neutral aluminium, calculations were performed in the EOL scheme \citep{Dyall} for 28 targeted states. These states belong to the $3s^{2}ns$ configurations with $n=4,5,6$, the $3s^{2}nd$ configurations with $n=3,...,6$, as well as the $3s3p^{2}$ and $3s^{2}5g$ configurations, characterized by even parity and on the other hand, the $3s^{2}np$ configurations with $n=3,..,6$, as well as the $3s^{2}4f$ and $3s^{2}5f$ configurations, characterized by odd parity. These configurations define the so-called multireference (MR). From initial calculations and analysis of the eigenvector compositions, we deduced that all $3p^2nl$ configurations, in addition to the targeted $3s^2nl$, give considerable contributions to the total wave functions and should be included in the MR. Following the active set (AS) approach \citep{Olsen1, Sturesson}, the CSF expansions (see eq. 2) were obtained by allowing single and restricted double (SD) substitutions of electrons from the reference (MR) orbitals to an AS of correlation orbitals. The AS is systematically increased by adding layers of orbitals to effectively build nearly complete wave functions. This is achieved by keeping track not only of the convergence of the computed excitation energies, but also of the other physical quantities of interest, such as the transition parameters here.

As a first step an MCDHF calculation was performed for the orbitals that are part of the MR. States with both even and odd parity were simultaneously optimized. Following this step, we continued to optimize six layers of correlation orbitals based on valence (VV) substitutions. The VV expansions were obtained by allowing SD substitutions from the three outer valence orbitals in the MR, with the restriction that there will be at most one substitution from orbitals with $n=3$. In this manner, the correlation orbitals will occupy the space between the inner $n=3$ valence orbitals and the outer orbitals involved in the higher Rydberg states (see Fig. 1). These orbitals have been shown to be of crucial importance for the transition probabilities, which are weighted towards this part of the space \citep{Asli,thesis}. The six correlation layers correspond to the $12s, 12p, 12d, 11f, 11g$ and $10h$ set of orbitals.

Each MCDHF calculation was followed by an RCI calculation for an extended expansion, obtained by single, double and triple (SDT) substitutions from the valence orbitals. As a final step, an RCI calculation was performed for the largest SDT valence expansion augmented by a core-valence (CV) expansion. The CV expansion was obtained by allowing SD substitutions from the valence orbitals and the $2p^6$ core, with the restriction that there will be at most one substitution from $2p^6$. All the RCI calculations included the Breit interaction and the leading QED effects. Accounting for CV correlation does not lower the total energies significantly, but it can have large effects on the energy separations and thus we considered it crucial. Core-valence correlation is also important for transition properties \citep{Hibbert}. Core-core (CC) correlation, obtained by allowing double excitations from the core, is known to be less important and has not been considered in the present work. The number of CSFs in the final even and odd state expansions, accounting for both VV and CV electron correlation, were 4 362 628 and 2 889 385, respectively, distributed over the different J symmetries.

\begin{figure}
	\centering
	\includegraphics[width=0.49\textwidth]{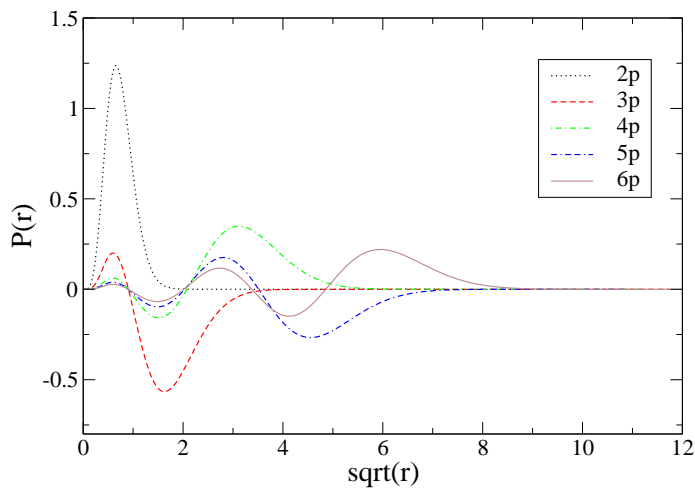}
	\caption{The \ion{Al}{I} Dirac-Fock radial orbitals for the $p$ symmetry, as a function of $\sqrt{r}$. The $2p$ orbital is part of the core and the orbitals from $n=3$ up to $n=6$ are part of the valence electron cloud. We note that these orbitals occupy different regions in space and the overlap between some of the Rydberg states will be minor.}
	\label{Fig1}
\end{figure}

\subsection{Al II}

In the singly ionized aluminium, the calculations were more extended, including 78 targeted states. These states belong to the $3s^2$ ground configuration, as well as the $3p^2$, the $3sns$ configurations with $n=4,...,7$, the $3snd$ with $n=3,...,6$, the $3s5g$ and $3s6g$ configurations, characterized by even parity and on the other hand, the $3snp$ configurations with $n=3,...,7$, the $3snf$ with $n=4,5,6$, as well as the $3s6h$ and $3p3d$ configurations, characterized by odd parity. These configurations define the multireference (MR). In the computations of \ion{Al}{II}, the EOL scheme was applied and the CSF expansions were obtained following the active set (AS) approach, accounting for VV and CV correlation. \ion{Al}{II} is less complex and the CSF expansions generated from (SD) substitutions are not as large as the ones in \ion{Al}{I}. Hence, we can afford both $2s$ and $2p$ orbitals to account for CV correlation. The $1s$ core orbital remained closed and as for \ion{Al}{I}, core-core correlation was neglected. The MCDHF calculations were performed in a similar way as the calculations in \ion{Al}{I}, yet no particular restrictions were imposed on the VV substitutions. We optimized six correlation layers corresponding to the $13s, 13p, 12d, 12f, 12g, 8h$ and $7i$ set of orbitals. Each MCDHF calculation was followed by an RCI calculation. At last, an RCI calculation was performed for the largest SD valence expansion augmented by the CV expansion. The number of CSFs in the final even and odd state expansions, accounting for both VV and CV electron correlation, were 911 795 and 1 269 797, respectively, distributed over the different J symmetries.

\section{Results}

\subsection{Al I}

\begin{table*}[h]
\caption{Computed excitation energies in cm$^{-1}$ for the 28 lowest states in \ion{Al}{I}. The energies are given as a function of the increasing active set of orbitals, accounting for VV correlation, where $n$ indicates the maximum principle quantum number of the orbitals included in the active set. In the third last column, the final energy values are displayed after accounting for CV correlation. The differences $\Delta E$ between the final computations and the observed values are shown in the last column. The sequence and labeling of the configurations and $LSJ$-levels are in accordance with the final (CV) computed energies. The $3s^24d~^2D$ term is assigned twice throughout the calculations (see also Table 2) and the subscripts $a$ and $b$ are being used to distinguish them. See text for more details.}
	\label{table:1}      
	\centering          
	\begin{tabular}{cccrrrrrrrrr}
		\hline
		\midrule
& & & \multicolumn{6}{c}{VV} & & & \\
Pos. & Conf. & $LSJ$ & $n=7$ & $n=8$ & $n=9$ & $n=10$ & $n=11$ & $n=12$ & CV & $E_\text{obs}$$^{(1)}$ & $\Delta E$ \\ 
\midrule

1 & $3s^2 \,3p~$     ~  &  $^2P_{  1/2  }^o$ &        0 &        0 &          0 &         0 &          0 &          0 &       0 &       0 & 0    \\ 
2 &                     &  $^2P_{  3/2  }^o$ &      108 &      108 &        108 &       108 &        108 &        108 &     104 &     112 & 8    \\
3 & $3s^2 \,4s~$       ~  &  $^2S_{  1/2  }$ &   25~318 &    25~377 &    25~416 &    25~419 &     25~427 &     25~429 &  25~196 &  25~348 & 152  \\
4 & $3s\,3p^2$         ~  &  $^4P_{  1/2  }$ &   27~788 &    27~966 &    28~073 &    28~085 &     28~109 &     28~111 &  28~863 &  29~020 & 157  \\
5 &                       &  $^4P_{  3/2  }$ &   27~833 &    28~011 &    28~118 &    28~130 &     28~154 &     28~156 &  28~907 &  29~067 & 160  \\
6 &                       &  $^4P_{  5/2  }$ &   27~906 &    28~085 &    28~191 &    28~204 &     28~227 &     28~230 &  28~981 &  29~143 & 162  \\
7 & $3s^2 \,3d~$       ~  &  $^2D_{  3/2  }$ &   32~211 &    32~077 &    32~135 &    32~139 &     32~150 &     32~150 &  32~414 &  32~435 & 21   \\
8 &                       &  $^2D_{  5/2  }$ &   32~212 &    32~079 &    32~137 &    32~141 &     32~152 &     32~152 &  32~416 &  32~437 & 21   \\
9 & $3s^2 \,4p~$     ~  &  $^2P_{  1/2  }^o$ &   32~770 &    32~879 &    32~935 &    32~937 &     32~946 &     32~949 &  32~801 &  32~950 & 149  \\
10 &                    &  $^2P_{  3/2  }^o$ &   32~786 &    32~894 &    32~951 &    32~952 &     32~962 &     32~964 &  32~814 &  32~966 & 152  \\
11 & $3s^2 \,5s~$    ~  &  $^2S_{  1/2  }$   &   37~493 &    37~637 &    37~693 &    37~694 &     37~704 &     37~706 &  37~512 &  37~689 & 177  \\
12 & $3s^2 \,4d~$    ~  &  $^2D_{  3/2~a}$   &   38~733 &    38~659 &    38~711 &    38~707 &     38~717 &     38~718 &  38~951 &  38~929 & -22  \\
13 &                    &  $^2D_{  5/2~a}$   &   38~736 &    38~664 &    38~717 &    38~712 &     38~722 &     38~724 &  38~957 &  38~934 & -23  \\
14 & $3s^2 \,5p~$    ~  &  $^2P_{  1/2  }^o$ &   40~038 &    40~187 &    40~252 &    40~249 &     40~259 &     40~262 &  40~101 &  40~272 & 171  \\
15 &                    &  $^2P_{  3/2  }^o$ &   40~043 &    40~193 &    40~258 &    40~255 &     40~265 &     40~268 &  40~106 &  40~278 & 172  \\
16 & $3s^2 \,4f~$    ~  &  $^2F_{  5/2  }^o$ &   41~050 &    41~209 &    41~282 &    41~287 &     41~297 &     41~300 &  41~163 &  41~319 & 156  \\
17 &                    &  $^2F_{  7/2  }^o$ &   41~050 &    41~209 &    41~282 &    41~287 &     41~297 &     41~300 &  41~163 &  41~319 & 156  \\ 
18 & $3s^2 \,6s~$    ~  &  $^2S_{  1/2  }$   &   41~897 &    42~069 &    42~133 &    42~135 &     42~144 &     42~143 &  41~964 &  42~144 & 180  \\
19 & $3s^2 \,4d~$    ~  &  $^2D_{  3/2~b}$   &   42~105 &    42~071 &    42~121 &    42~108 &     42~119 &     42~121 &  42~232 &  42~234 & 2    \\
20 &                    &  $^2D_{  5/2~b}$   &   42~109 &    42~075 &    42~126 &    42~112 &     42~123 &     42~125 &  42~237 &  42~238 & 1    \\   
21 & $3s^2 \,6p~$    ~  &  $^2P_{  1/2  }^o$ &   43~076 &    43~246 &    43~316 &    43~311 &     43~321 &     43~324 &  43~160 &  43~335 & 175  \\
22 &                    &  $^2P_{  3/2  }^o$ &   43~079 &    43~249 &    43~318 &    43~313 &     43~324 &     43~326 &  43~162 &  43~338 & 176  \\
23 & $3s^2 \,5f~$    ~  &  $^2F_{  5/2  }^o$ &   43~549 &    43~721 &    43~795 &    43~801 &     43~811 &     43~813 &  43~660 &  43~831 & 171  \\
24 &                    &  $^2F_{  7/2  }^o$ &   43~549 &    43~721 &    43~795 &    43~801 &     43~811 &     43~813 &  43~660 &  43~831 & 171  \\
25 & $3s^2 \,5g~$    ~  &  $^2G_{  7/2  }$   &   43~576 &    43~763 &    43~838 &    43~845 &     43~856 &     43~859 &  43~687 &  43~876 & 189  \\
26 &                    &  $^2G_{  9/2  }$   &   43~576 &    43~763 &    43~838 &    43~845 &     43~856 &     43~859 &  43~687 &  43~876 & 189  \\
27 & $3s^2 \,5d~$    ~  &  $^2D_{  3/2  }$   &   44~034 &    44~059 &    44~115 &    44~096 &     44~106 &     44~109 &  44~126 &  44~166 & 40   \\
28 &                    &  $^2D_{  5/2  }$   &   44~036 &    44~062 &    44~117 &    44~099 &     44~109 &     44~111 &  44~129 &  44~169 & 40   \\

\hline	
\end{tabular}
\tablebib{ $^{(1)}$NIST Atomic Spectra Database 2018 \citep{Kramida}.}
\end{table*}

\begin{table*}[h]
\caption{$LS$-composition of the computed states belonging to the strongly mixed $3s^2nd$ Rydberg series in \ion{Al}{I}. The three most dominant $LS$-components are displayed. The first percentage value corresponds to the assigned configuration and term. In all these cases, the percentages for the two different $LSJ$-levels are the same and therefore given in the same line. In the last column, we provide the labeling of the corresponding observed terms as given in the NIST Database. The first column refers to the positions according to Table 1.}
\label{table:2}      
\centering          
\begin{tabular}{ccccc}
\hline
\midrule
Pos. & Conf. & $LSJ$ & $LS$-composition & Label used in NIST \\ 		
\midrule		
7,8   & $3s^{2}\,3d~$ & $^{2}D_{3/2,5/2}$  & 0.67 + 0.19~$3s\,3p^{2}~^{2}D$ + 0.04~$3s^{2}\,4d~^{2}D$ & $3s^{2}\,3d~~~~^{2}D_{3/2,5/2}$ \\ 
\\
12,13  & $3s^{2}\,4d~$ & $^{2}D_{3/2,5/2}$$_{~a}$ & 0.41 + 0.22~$3s^{2}\,3d~^{2}D$ + 0.21~$3s\,3p^{2}~^{2}D$ & $3s^{2}\,nd~~y~^{2}D_{3/2,5/2}$ \\  
\\
19,20  & $3s^{2}\,4d~$ & $^{2}D_{3/2,5/2}$$_{~b}$ & 0.44 + 0.25~$3s^{2}\,5d~^{2}D$ + 0.15~$3s\,3p^{2}~^{2}D$ & $3s^{2}\,4d~~~~^{2}D_{3/2,5/2}$ \\ 
\\
27,28  & $3s^{2}\,5d~$ & $^{2}D_{3/2,5/2}$  & 0.58 + 0.19~$3s^{2}\,6d~^{2}D$ + 0.10~$3s\,3p^{2}~^{2}D$ & $3s^{2}\,5d~~~~^{2}D_{3/2,5/2}$ \\ 	
\midrule
\end{tabular}
\tablebib{ $^{(1)}$\citet{Kramida}. }
\end{table*}

\begin{table*}[h]
\caption{Observed and computed excitation energies in cm$^{-1}$ for the 10 and 20 lowest states in \ion{Al}{I} and \ion{Al}{II}, respectively. In the last two columns, the difference $\Delta E$ between observed and computed energies is compared for the current RCI and previous MCHF-BP calculations.}
\label{table:3}      
\centering          
\begin{tabular}{cccrrrc}
\hline
\midrule
Pos. &  Conf.  &  $LSJ$  & $E_\text{obs}$$^{(1)}$ & $E_\text{RCI}$$^{(2)}$ & $\Delta  E_\text{RCI}$$^{(2)}$ & $\Delta E_\text{MCHF-BP}$$^{(3)}$  \\
\midrule
\multicolumn{7}{c}{\ion{Al}{I}} \\
\midrule
1  & $3s^2 \,3p~$ ~  &  $^2P_{  1/2  }^o$  &      0 &        0   &    0   &      0    \\
2  &                 &  $^2P_{  3/2  }^o$  &    112 &      104   &	  8   &     22    \\
3  & $3s^2 \,4s~$ ~  &  $^2S_{  1/2  }$    & 25~348 &   25~196   &	152	  &   -235    \\
4  & $3s\,3p^2$   ~  &  $^4P_{  1/2  }$    & 29~020 &   28~863   &	157	  &    940    \\
5  &                 &  $^4P_{  3/2  }$    & 29~067 &   28~907   &	160	  &    949    \\
6  &                 &  $^4P_{  5/2  }$    & 29~143 &   28~981   &	162	  &    964    \\
7  & $3s^2 \,3d~$ ~  &  $^2D_{  3/2  }$    & 32~435 &   32~414   &	 21   &    250    \\
8  &                 &  $^2D_{  5/2  }$    & 32~437 &   32~416   &	 21   &    251    \\
9  & $3s^2 \,4p~$ ~  &  $^2P_{  1/2  }^o$  & 32~950 &   32~801   &	149	  &    -98    \\
10 &                 &  $^2P_{  3/2  }^o$  & 32~966 &   32~814   &	152	  &    -94    \\	
\midrule
\multicolumn{7}{c}{\ion{Al}{II}} \\
\midrule	
1  &  $3s^2 ~$    &  $^1S_{   0   } $   &      0    &  0       &     0   &     0    \\
2  &  $3s\,3p~$   &  $^3P_{   0   }^o$  &  37~393   &  37~445  &   -52   &     9    \\
3  &              &  $^3P_{   1   }^o$  &  37~454   &  37~503  &   -49   &     8    \\
4  &              &  $^3P_{   2   }^o$  &  37~578   &  37~626  &   -48   &     6    \\
5  &              &  $^1P_{   1   }^o$  &  59~852   &  59~982  &  -130   &  -177    \\
6  &  $3p^2 ~$    &  $^1D_{   2   } $   &  85~481   &  85~692  &  -211   &  -305    \\
7  &  $3s\,4s~$   &  $^3S_{   1   } $   &  91~275   &  91~425  &  -150   &  -376    \\
8  &  $3p^2 ~$    &  $^3P_{   0   } $   &  94~085   &  94~211  &  -126   &  -107    \\
9  &              &  $^3P_{   1   } $   &  94~147   &  94~264  &  -117   &  -111    \\
10 &              &  $^3P_{   2   } $   &  94~269   &  94~375  &  -106   &  -113    \\
11 &  $3s\,4s~$   &  $^1S_{   0   } $   &  95~351   &  95~543  &  -192   &  -400    \\
12 &  $3s\,3d~$   &  $^3D_{   2   } $   &  95~549   &  95~791  &  -242   &  -527    \\
13 &              &  $^3D_{   1   } $   &  95~551   &  95~794  &  -243   &  -527    \\
14 &          	  &  $^3D_{   3   } $   &  95~551   &  95~804  &  -253   &  -529    \\
15 &  $3s\,4p~$   &  $^3P_{   0   }^o$  & 105~428   & 105~582  &  -154   &  -357    \\
16 &              &  $^3P_{   1   }^o$  & 105~442   & 105~594  &  -152   &  -360    \\
17 &              &  $^3P_{   2   }^o$  & 105~471   & 105~623  &  -152   &  -363    \\
18 &              &  $^1P_{   1   }^o$  & 106~921   & 107~132  &  -211   &  -365    \\
19 &  $3s\,3d~$   &  $^1D_{   2   } $   & 110~090   & 110~330  &  -240   &  -475    \\
20 &  $3p^2 ~$    &  $^1S_{   0   } $   & 111~637   & 112~086  &  -449   &  -445    \\
\midrule
\end{tabular}
\tablebib{ $^{(1)}$\citet{Kramida}; $^{(2)}$present calculations; $^{(3)}$\citet{Froese}.}
\end{table*}

In Table 1, the computed excitation energies, based on VV correlation, are given as a function of the increasing active set of orbitals. After adding the $n=11$ correlation layer, we note that the energy values for all 28 targeted states have converged. For comparison, in the second last column the observed energies from the NIST Atomic Spectra Database \citep{Kramida} are displayed. All energies but the ones belonging to the $3s3p^2$ configuration are already in good agreement with the NIST recommended values. The relative differences between theory and experiment for all three levels of the quartet $3s3p^2$ $^4P$ state is $3.1\%$, while the mean relative difference for the rest of the states is less than $0.2\%$. In the third last column, the computed excitation energies after accounting for CV correlation are displayed. When taking into account CV effects the agreement with the observed values is better overall. But most importantly, for the $3s3p^2~^4P$ levels the relative differences between observed and computed values decrease to less than $0.6\%$. The likelihood of the $1s^22s^22p^6$ core to overlap with the $3s3p^2$ cloud of electrons is much less relatively to the one for $3s^2nl$. Consequently, when CV correlation is taken into account the lowering of the $3s3p^2$ energy levels is much smaller than for levels belonging to any $3s^2nl$ configuration. Thus, the adjustments to the separation energies will be minor between the ground state $3s^23p$ and $3s^2nl$ levels, but significant between the $3s^23p$ and $3s3p^2$ levels. In the last column of Table 1, the differences $\Delta E$$=E_\text{obs}-E_\text{theor}$, between the final (CV) computed and the observed energies, are also displayed. In principle, there are two groups of values, with the one consisting of the $3s^2nd$ configurations exhibiting the smallest absolute discrepancies from the observed energies. For the rest, the absolute discrepancies are somewhat larger. 
 
In the calculations, the labeling of the eigenstates is determined by the CSF with the largest coefficient in the expansion of eq. 2. When the same label is assigned to different eigenstates, a detailed analysis can be performed by displaying their $LS$-compositions. In Table 1, we note that two of the states have been assigned the same label, i.e. $3s^24d ~^2D$, and thus the subscripts $a$ and $b$ are used to distinguish them. In Table 2, we give the $LS$-composition of all computed $3s^2nd~^2D$ states, including the three most dominant CSFs. The $3s^24d~^2D$ term appears twice as the CSF with the largest $LS$-composition. Moreover, the admixture of the $3s3p^2~^2D$ in the lowest four $3s^2nd~^2D$ states is rather strong and adds up to $65\%$. That being so, the $3s3p^2~^2D$ does not exist in the calculated spectrum as a localized state. For comparison, in the last column of Table 2, the labeling of the observed $3s^2nd~^2D$ states is also given. In the observed configurations presented by NIST \citep{Kramida}, the second highest $3s^2nd~^2D$ term has not been given any specific label and it is therefore designated as $y~^2D$. The higher $^2D$ terms are designated as $3s^24d,~3s^25d$ and so on.

In Table 3, the current results for the lowest excitation energies are compared with the ones from the MCHF-BP calculations by \citet{Froese}. The latter calculations are extended up to levels corresponding to the doublet $3s^24p~^2P$ state. The differences $\Delta E$ between observed and computed energies are given in the last two columns for the different computational approaches. As seen, when using the current MCDHF and RCI method, the agreement with the observed energies is substantially improved for all levels and in particular, for the ones belonging to the quartet $3s3p^2~^4P$ state. In the MCHF-BP calculations, core-valence correlation was neglected. As mentioned above and also acknowledged by \citet{Froese}, capturing such correlation effects is crucial for 3s-hole states, such as states with significant $3s3p^2$ composition. Furthermore, the $\Delta E_\text{MCHF-BP}$ values do not always have the same sign, while the $\Delta E_\text{RCI}$ differences are consistently positive. This is particularly important when calculating transition properties. On average, properties for transitions between two levels for which the differences $\Delta E_\text{MCHF-BP}$ have opposite sign will be estimated less accurately.

\begin{table*}[h]
\caption{Statistical analysis of the uncertainties of the computed transition rates in \ion{Al}{I} and \ion{Al}{II}. The transition rates are arranged in four groups and in the forth column, the number of transitions belonging to each group is given. In the last three columns, the average value, the value $Q_3$ containing $75\%$ of the lowest computed $dT$ values, as well as the maximum value are given for each group of transitions. All transition rates are given in s$^{-1}$.}
\label{table:4}      
\centering          
\begin{tabular}{ccccccc}
\hline
\midrule
 &  $A^\text{low}_\text{RCI}$  & $A^\text{high}_\text{RCI}$  & No.Trans. & $\langle dT \rangle$ & $Q_3$ & max \\
\midrule
\multicolumn{7}{c}{\ion{Al}{I}} \\
\midrule
1  &            & 1.00E+05  & 31  & 0.62  & 0.83  & 0.98 \\  
2  &  1.00E+05  & 1.00E+06  & 25  & 0.29  & 0.37  & 0.81 \\
3  &  1.00E+06  & 1.00E+07  & 24  & 0.055 & 0.076 & 0.15 \\
4  &  1.00E+07  &           & 20  & 0.043 & 0.073 & 0.14 \\
\midrule
\multicolumn{7}{c}{\ion{Al}{II}} \\
\midrule
1  &            & 1.00E+05  & 109 & 0.07  & 0.11  & 0.61 \\
2  & 1.00E+05   & 1.00E+06  &  81 & 0.09  & 0.11  & 0.67 \\
3  & 1.00E+06   & 1.00E+07  &  99 & 0.043 & 0.036 & 0.39 \\
4  & 1.00E+07   &           & 141 & 0.011 & 0.009 & 0.12 \\		
\midrule
\end{tabular}
\end{table*}

The complete transition data, for all computed E1 transitions in \ion{Al}{I}, can be found in Table 10. In Table 10, the transition energies, wavelengths and the length form of the transition rates $A$ and weighted oscillator strengths $gf$ are given. Based on the agreement between the length and velocity forms of the computed transition rates $A_\text{RCI}$, a statistical analysis of the uncertainties can be preformed. The transitions were arranged in four groups based on the magnitude of the $A_\text{RCI}$ values. The first two groups contain all the weak transitions with transition rates up to $A=10^6$ s$^{-1}$, while the next two groups contain the strong transitions with $A>10^6$ s$^{-1}$. In Table 4, the average value of the uncertainties $\langle dT \rangle$ (see eq. 4) is given for each group of transitions. To better understand how the individual uncertainties $dT$ are distributed, the maximum value, as well as the value $Q_3$ containing $75\%$ of the lowest computed $dT$ values (third quartile) are also given in Table 4. When examining the predicted uncertainties of the individual groups, we deduce that for all the strong transitions $dT$ always remains below $15\%$. In fact, the majority of the strong transitions is associated with uncertainties of the order of a few per cent, which justifies the low average values.  Contrary to the strong transitions, the weaker transitions are associated with considerably larger uncertainties. This is even more pronounced for the first group of transitions, where $A$ is less than $10^5$ s$^{-1}$. The weak E1 transitions are challenging, and therefore interesting, from a theoretical point of view, yet they are less likely to be observed. The computation of transition properties in the system of \ion{Al}{I} is overall far from trivial due to the extreme mixing of the $3s^2nd~^2D$ series. Transitions involving any $^2D$ state as upper or lower level appear to be associated with large uncertainties. However, the predicted energy separations are in excellent agreement with observations, meaning that the $LS$-composition of the $3s^2nd~^2D$ states is well described. This fact should serve as a quality indicator of the computed transition data.

%-------------------------------TABLE 5--------------------------------------------------------
\begin{table*}
\caption[]{Comparison between computed and observed transition rates $A$ in s$^{-1}$, for selected transitions in \ion{Al}{I}. The present values from the RCI calculations are given in the third column. In the next two columns, theoretical values from former MCHF-BP and close coupling (CC) calculations are displayed. The CC values complement the MCHF-BP ones, which are restricted to transitions between levels in the lower part of the \ion{Al}{I} spectrum. The last two columns contain the results from experimental observations. The experimental results go along with a letter-grade, whenever accessible, which indicates their accuracy level.}
\label{table:5}      
\centering          
\begin{tabular}{lllllll}
\hline
\midrule
Upper & Lower &  $A_\text{RCI}$$^{(2)}$ &  $A_\text{MCHF-BP}$$^{(3)}$ & $A_\text{CC}$$^{(4a)}$ & $A_\text{obs}$$^{(5),(6),(7)}$ & $A_\text{obs}$$^{(8)}$  \\ 
\midrule
$3s^2 \,4s~^2S_{ 1/2  }$    & $3s^2 \,3p~^2P_{ 1/2  }^o$    &   4.966E+07   & 5.098E+07  &  & 4.93E+07$^{~(5)}$$~^{C}$& 4.70E+07 $^{B}$ \\
                            & $3s^2 \,3p~^2P_{ 3/2  }^o$    &   9.884E+07   & 10.15E+07  &  & 9.80E+07$^{~(5)}$$~^{C}$ & 9.90E+07 $^{B}$ \\
$3s^2 \,5s~^2S_{ 1/2  }$  & $3s^2 \,3p~^2P_{ 1/2  }^o$   &   1.277E+07   &  &  & 1.33E+07$^{~(5)}$$~^{C}$& 1.42E+07 $^{C+}$ \\
                          & $3s^2 \,3p~^2P_{ 3/2  }^o$   &   2.534E+07   &  &  & 2.64E+07$^{~(5)}$$~^{C}$& 2.84E+07 $^{C+}$ \\

$3s^2 \,5s~^2S_{ 1/2  }$   & $3s^2 \,4p~^2P_{ 1/2  }^o$   &   3.815E+06   &  &  &  & 3.00E+06 $^{D}$ \\
                           & $3s^2 \,4p~^2P_{ 3/2  }^o$   &   7.599E+06   &  &  &  & 6.00E+06 $^{D}$ \\

$3s^2 \,4p~^2P_{ 1/2  }^o$ & $3s^2 \,4s~^2S_{ 1/2  }$     &   1.580E+07   & 1.507E+07 &  & 1.69E+07$^{~(6)}$ $^{C+}$ & 1.60E+07 $^{A}$ \\
$3s^2 \,4p~^2P_{ 3/2  }^o$ & $3s^2 \,4s~^2S_{ 1/2  }$     &   1.587E+07   & 1.514E+07 &  & 1.69E+07$^{~(6)}$ $^{C+}$ & 1.50E+07 $^{B}$ \\

$3s^2 \,3d~^2D_{ 3/2  }$   & $3s^2 \,3p~^2P_{ 1/2  }^o$   &   6.542E+07   & 5.651E+07  &  & 6.30E+07$^{~(5)}$ $^{C}$ & 5.90E+07 $^{C+}$ \\
                           & $3s^2 \,3p~^2P_{ 3/2  }^o$   &   1.321E+07   & 1.140E+07  &  &             & (1.20)E+07 \\
$3s^2 \,3d~^2D_{ 5/2  }$   & $3s^2 \,3p~^2P_{ 3/2  }^o$   &   7.877E+07   & 6.806E+07  &  & 7.40E+07$^{~(5)}$ $^{C}$ & 7.10E+07 $^{A}$ \\

$3s^2 \,4d~^2D_{ 3/2~a}$   & $3s^2 \,3p~^2P_{ 1/2  }^o$   &   1.722E+07   &  &          & 1.92E+07$^{~(7)}$ $^{C+}$ &  \\
                           &                              &               &  &          & 2.30E+07$^{~(5)}$ $^{C}$ &   \\
                           & $3s^2 \,3p~^2P_{ 3/2  }^o$   &   3.293E+06   &  & 5.99E+06 & 3.80E+06$^{~(7)}$ $^{C+}$ &  \\
                           &                              &               &  &          & 4.40E+06$^{~(5)}$ $^{C}$ &   \\ 
$3s^2 \,4d~^2D_{ 5/2~a}$   & $3s^2 \,3p~^2P_{ 3/2  }^o$   &   2.010E+07   &  & 3.60E+07 & 2.30E+07$^{~(7)}$ &  \\
                           &                              &               &  &          & 2.80E+07$^{~(5)}$ $^{C}$ &   \\

$3s^2 \,4d~^2D_{ 3/2~b}$   & $3s^2 \,3p~^2P_{ 1/2  }^o$   &   7.128E+07   &  & 7.61E+07 & 7.20E+07$^{~(5)}$ $^{C}$ &  \\
                           &                              &               &  &          & 5.26E+07$^{~(7)}$ &   \\
                           & $3s^2 \,3p~^2P_{ 3/2  }^o$   &   1.386E+07   &  & 1.51E+07 & 1.40E+07$^{~(5)}$ $^{C}$ &  \\
                           &                              &               &  &          & 1.05E+07$^{~(7)}$ $^{A}$ &   \\
$3s^2 \,4d~^2D_{ 5/2~b}$   & $3s^2 \,3p~^2P_{ 3/2  }^o$   &   8.412E+07   &  & 9.07E+07 & 8.60E+07$^{~(5)}$ $^{C}$ &  \\
                           &                              &               &  &          & 6.31E+07$^{~(7)}$ &   \\

$3s^2 \,5d~^2D_{ 3/2  }$   & $3s^2 \,3p~^2P_{ 1/2  }^o$   &   8.204E+07   &  &         & 6.60E+07$^{~(5)}$ $^{C}$ &  \\
                           &                              &               &  &         & 5.76E+07$^{~(7)}$ &   \\
                           & $3s^2 \,3p~^2P_{ 3/2  }^o$   &   1.596E+07   &  & 1.26E+07 & 1.30E+07$^{~(5)}$ $^{C}$ &  \\
                           &                              &               &  &          & 1.15E+07$^{~(7)}$ &   \\
$3s^2 \,5d~^2D_{ 5/2  }$   & $3s^2 \,3p~^2P_{ 3/2  }^o$   &   9.706E+07   &  & 7.58E+07 & 7.90E+07$^{~(5)}$ $^{C}$ &  \\
                           &                              &               &  &          & 6.91E+07$^{~(7)}$ &   \\
\midrule	
\end{tabular}
\tablefoot{All theoretical transition rates are presented in length form. The correspondance between the accuracy ratings and the estimated relative uncertainty of the experimental results is: $A:\le3\%,~B:\le10\%,~C+:\le15\%,~C:\le25\%,~D+:\le30\%,~D:\le50\%$}
\tablebib{ $^{(2)}$Present calculations; $^{(3)}$\citet{Froese}; $^{(4a)}$\citet{Mendoza}; $^{(5)}$\citet{WieseMartin}; $^{(6)}$\citet{Buurman}; $^{(7)}$\citet{Davidson}; $^{(8)}$\citet{Vujnovic}.}
\end{table*}

Transition rates $A_\text{obs}$ evaluated from experimental measurements are, in Table 5, compared with the current RCI theoretical values, as well as values from the MCHF-BP calculations by \citet{Froese} and the close coupling (CC) calculations by \citet{Mendoza}. Despite the fact that the measurements by \citet{Davidson} are more recent compared to the compiled values by \citet{WieseMartin}, the latter seem to be in better overall agreement with the transition rates predicted by the RCI calculations. In all cases, the $A_\text{RCI}$ values fall into the range of the estimated uncertainties by \citet{WieseMartin}. The only exceptions are the transitions with $3s^24d~^2D_{3/2,5/2~a}$ as upper levels, for which the $A_\text{RCI}$ values agree better with the ones suggested by \citet{Davidson}. Although the evaluated transition rates by \citet{Vujnovic} slightly differ from the other observations, they are still in fairly good agreement with the present work. For the $3s^24p~^2P^o_{3/2}~\rightarrow~3s^24s~^2S_{1/2}$ and $3s^23d~^2D_{5/2}~\rightarrow~3s^23p^2P^o_{3/2}$ transitions, the values by \citet{Vujnovic} are better reproduced by the $A_\text{MCHF-BP}$ results, yet not enough correlation is included in the calculations by \citet{Froese} and the transition rates predicted by the RCI calculations should be considered as more accurate. Whenever values from the close coupling (CC) calculations are presented to complement the MCHF-BP results, the $A_\text{RCI}$ values appear to be in better agreement with the experimental values. Exceptionally, for the $3s^25d~^2D_{3/2,5/2}~\rightarrow~3s^23p~^2P^o_{3/2}$ transitions, the $A_\text{CC}$ values by \citet{Mendoza} approach more the corresponding experimental values. Even so, the $A_\text{RCI}$ values are still within the given experimental uncertainties. One should bear in mind that according to the estimation of uncertainties by \citet{Kelleherb} the $A_\text{CC}$ values carry relative uncertainties up to $30\%$. On the contrary, based on the agreement between length and velocity forms, the estimated uncertainties of the current RCI calculations for the above mentioned transitions are of the order of a few percent. Therefore, we suggest that the current transition rates are used as a reference.

From the computed E1 transition rates, the lifetimes of the excited states are estimated. Transition data for other transitions than E1 have not been computed in this work, since the contributions to the lifetimes from magnetic or higher electric multipoles are expected to be negligible. In Table 6, the currently computed lifetimes are given in both length $\tau_l$ and velocity $\tau_\upsilon$ forms. The agreement between these two forms probes the level of accuracy of the calculations. Because of the poor agreement between the length and velocity form of the quartet $3s3p^2~^4P$ and doublet $3s^26p~^2P$ states, the average relative difference appears overall to be $\sim8\%$. The differences between the length and velocity gauges of the quartet $3s3p^2~^4P$ states are of the order of $25\%$ in average. These long-lived states are associated with weak transitions and computations involving such transitions are, as mentioned before, rather challenging. In addition, we note that the relative differences corresponding to the $3s^26p~^2P$ states exceed $40\%$. As the computations involve Rydberg series, states between the lowest and highest computed levels might not occupy the same region in space. Nevertheless, these states are part of the same multireference (MR). The highest computed levels correspond to configurations with orbitals up to $n=6$, such as $3s^26p$. To obtain a better description of those levels it is probably necessary to perform initial calculations including in the MR $3s^2nl$ configurations with $n=7$ and maybe even $n=8$. This would lead to a more complete and balanced orbital set \citep{Asli}. When excluding the above mentioned states, the mean relative difference between $\tau_l$ and $\tau_\upsilon$ is $\sim3\%$, which is satisfactory. 

In Table 6, the lifetimes from the current RCI calculations are also compared with results from the MCHF-BP calculations by \citet{Froese} and observations. Only for the $3s^24p~^2P$ state, separated observed values of the lifetimes are given for the two fine-structure components. For the rest of the measured lifetimes, a single value for the two fine-structure levels is provided. As seen, the overall agreement between the theoretical and the measured lifetimes $\tau_\text{obs}$ is rather good. However, the measured lifetimes are better represented by the current RCI results compared to the MCHF-BP ones. For most of the states, the differences between the RCI and MCHF-BP values are small, except for the levels of the quartet $3s3p^2~^4P$ state. For these long-lived states, no experimental lifetimes exist for comparison.

\begin{table*}
\caption{Comparison between computed and observed lifetimes $\tau$ in seconds, for the 26 lowest excited states in \ion{Al}{I}. For the current RCI calculations both length $\tau_l$ and velocity $\tau_\upsilon$ forms are displayed. In the second last column, the predicted lifetimes from MCHF-BP calculations are given in length form. The last column contains available lifetimes from experimental measurements, together with their uncertainties.}
\label{table:6}      
\centering          
\begin{tabular}{clccccc}
\hline
\midrule
& & & \multicolumn{2}{c}{RCI $^{(2)}$} & MCHF-BP $^{(3)}$ & Expt.$^{(6),(7),(9)}$ \\ 
Pos. & Conf. & $LSJ$ & $\tau_l$ & $\tau_\upsilon$ & $\tau_l$ & $\tau_\text{obs}$ \\ 
\midrule
1  & $3s^2 \,4s~$ & $^2S_{ 1/2 }$             &   6.734E-09 &  6.745E-09  &  6.558E-09 &  6.85(6)E-09 $^{(6)}$ \\
2  & $3s\,3p^2 $  & $^4P_{ 1/2 }$             &   1.652E-03 &  1.182E-03  &  4.950E-03 &                \\
3  &                         & $^4P_{ 3/2 }$  &   6.702E-03 &  6.911E-03  &  13.24E-03 &                \\
4  &                         & $^4P_{ 5/2 }$  &   2.604E-03 &  3.681E-03  &  9.486E-03 &                \\
5  & $3s^2 \,3d~$ & $^2D_{ 3/2 }$             &   1.272E-08 &  1.372E-08  &  1.472E-08 &  1.40(2)E-08 $^{(6)}$ \\
6  &              & $^2D_{ 5/2 }$             &   1.270E-08 &  1.371E-08  &  1.469E-08 &  1.40(2)E-08 $^{(6)}$ \\
7  & $3s^2 \,4p~$ & $^2P_{ 1/2 }^o$           &   6.329E-08 &  6.357E-08  &  6.621E-08 &  6.05(9)E-08 $^{(9)}$ \\
8  &              & $^2P_{ 3/2 }^o$           &   6.300E-08 &  6.328E-08  &  6.590E-08 &  6.5~(2)~E-08 $^{(9)}$ \\
9  & $3s^2 \,5s~$ & $^2S_{ 1/2 }$             &   2.019E-08 &  2.027E-08  &            &  1.98(5)E-08 $^{(6)}$ \\
10 & $3s^2 \,4d~$ & $^2D_{ 3/2~a}$            &   3.117E-08 &  2.919E-08  &            &  2.95(7)E-08 $^{(6)}$ \\
11 &              & $^2D_{ 5/2~a}$            &   3.158E-08 &  2.953E-08  &            &  2.95(7)E-08 $^{(6)}$ \\
12 & $3s^2 \,5p~$ & $^2P_{ 1/2 }^o$           &   2.448E-07 &  2.532E-07  &            &  2.75(8)E-07 $^{(6)}$ \\
13 &              & $^2P_{ 3/2 }^o$           &   2.429E-07 &  2.512E-07  &            &  2.75(8)E-07 $^{(6)}$ \\
14 & $3s^2 \,4f~$ & $^2F_{ 5/2 }^o$           &   6.041E-08 &  6.162E-08  &            &               \\
15 &              & $^2F_{ 7/2 }^o$           &   6.041E-08 &  6.160E-08  &            &               \\
16 & $3s^2 \,6s~$ & $^2S_{ 1/2 }$             &   4.812E-08 &  4.885E-08  &            &               \\
17 & $3s^2 \,4d~$ & $^2D_{ 3/2~b}$            &   1.136E-08 &  1.083E-08  &            &  1.32(3)E-08 $^{(7)}$ \\
18 &              & $^2D_{ 5/2~b}$            &   1.150E-08 &  1.093E-08  &            &  1.32(3)E-08 $^{(7)}$ \\
19 & $3s^2 \,6p~$ & $^2P_{ 1/2 }^o$           &   4.886E-07 &  6.952E-07  &            &               \\
20 &              & $^2P_{ 3/2 }^o$           &   4.845E-07 &  6.882E-07  &            &               \\
21 & $3s^2 \,5f~$ & $^2F_{ 7/2 }^o$           &   1.176E-07 &  1.172E-07  &            &               \\
22 &              & $^2F_{ 5/2 }^o$           &   1.175E-07 &  1.172E-07  &            &               \\
23 & $3s^2 \,5g~$ & $^2G_{ 7/2 }$             &   2.301E-07 &  2.315E-07  &            &               \\
24 &              & $^2G_{ 9/2 }$             &   2.301E-07 &  2.315E-07  &            &               \\
25 & $3s^2 \,5d~$ & $^2D_{ 3/2 }$             &   1.011E-08 &  9.855E-09  &            &  14.0(2)E-09 $^{(7)}$ \\
26 &              & $^2D_{ 5/2 }$             &   1.020E-08 &  9.921E-09  &            &  14.0(2)E-09 $^{(7)}$ \\
\midrule	

\end{tabular}
\tablebib{ $^{(2)}$Present calculations; $^{(3)}$\citet{Froese}; $^{(6)}$\citet{Buurman}; $^{(7)}$\citet{Davidson}; $^{(9)}$\citet{BuurDonsz}. }
\end{table*}

\subsection{Al II}

In Table 8, the computed excitation energies, based on VV correlation, are given as a function of the increasing active set of orbitals. When adding the $n=12$ correlation layer, the values for all computed energy separations have converged. The agreement with the NIST \citep{Kramida} observed energies is, at this point, fairly good. The mean relative difference between theory and experiment is of the order of $1.2\%$. However, when accounting for CV correlation the agreement with the observed values is significantly improved, resulting in a mean relative difference being less than $0.2\%$. Accounting for CV effects also results in a labeling of the eigenstates that matches with observations. For instance, when only VV correlation is taken into account, the $^3F$ triplet with the highest energy is labeled as a $3s6f$ level. After taking CV effects into account, the eigenstates of this triplet are assigned the $3p3d$ configuration being now the one with the largest expansion coefficient, which agree with observations. There are no experimental excitation energies for the singlet and triplet $3s6h~^{1,3}H$ terms. In the last column of Table 8, the differences $\Delta E$, between computed and observed energies, are displayed. All $\Delta E$ values maintain the same sign, being negative.  

In Table 3, a comparison between the present computed excitation energies and the ones from the MCHF-BP calculations by \citet{Froese} is also performed for \ion{Al}{II}. The latter spectrum calculations are extended up to levels corresponding to the singlet $3p^2~^1S$ state and all types of correlation, i.e. VV, CV and CC, were accounted for. Both computational approaches are highly accurate, yet the majority of the levels is better represented by the current RCI results. The average relative difference for the RCI values is $\sim0.2\%$ and for the MCHF-BP $\sim0.3\%$. Moreover, the $\Delta E_\text{MCHF-BP}$ values do not always maintain the same sign, while the $\Delta E_\text{RCI}$ values do. Hence, in general, the MCHF-BP calculations do not predict the transition energies and properties as precisely as the present RCI method.

\begin{table*}
\caption[]{Comparison between computed and observed transition rates $A$ in s$^{-1}$, for selected transitions in \ion{Al}{II}. The present values from the RCI calculations are given in the third column. In the next two columns, theoretical values from former MCHF-BP, close coupling (CC), configuration interaction (CI) and B-spline configuration interaction (BSCI) calculations are displayed. The last column contains the results from experimental observations. All theoretical transition rates are presented in length form.}
\label{table:6}      
\centering          
\begin{tabular}{llllll}
\hline
\midrule
Upper & Lower & $A_\text{RCI}$$^{(2)}$ &  $A_\text{theor}$$^{(3),(4b)}$ & $A_\text{theor}$$^{(10),(11)}$ & $A_\text{obs}$$^{(8)}$\\ 
\midrule
$3s3p~^3P_{  1  }^o$ &   $3s^2~^1S_{  0  }$     & 3.054E+03  &     3.277E+03$^{~(3)}$  &                     & 3.30E+03$^{~(13)}$  \\
$3s3p~^1P_{  1  }^o$ &   $3s^2~^1S_{  0  }$     & 1.404E+09  &     1.400E+09$^{~(3)}$  & 1.486E+09$^{~(11)}$ & 1.45E+09$^{~(12)}$  \\
$3s4s~^3S_{  1  }$   &   $3s3p~^3P_{  0  }^o$   & 8.612E+07  &     8.572E+07$^{~(3)}$  &                     &                     \\
&   $3s3p~^3P_{  1  }^o$   & 2.555E+08  &     2.547E+08$^{~(3)}$  &                     &                     \\
&   $3s3p~^3P_{  2  }^o$   & 4.173E+08  &     4.162E+08$^{~(3)}$  &                     &                     \\
$3s4s~^1S_{  0  }$   &   $3s3p~^1P_{  1  }^o$   & 3.422E+08  &     3.455E+08$^{~(3)}$  & 3.408E+08$^{~(11)}$ &                     \\
$3p^2~^1D_{  2  }$   &   $3s3p~^1P_{  1  }^o$   & 2.523E+05  &     3.804E+05$^{~(3)}$  & 3.980E+05$^{~(11)}$ & 1.84E+04$^{~(8)}$   \\
&   $3s3p~^3P_{  1  }^o$   & 1.790E+04  &     2.016E+04$^{~(3)}$  &                     & 0.19E+04$^{~(8)}$   \\
&   $3s3p~^3P_{  2  }^o$   & 2.827E+04  &     3.141E+04$^{~(3)}$  &                     & 0.30E+04$^{~(8)}$   \\
$3p^2~^3P_{  0  }$   &   $3s3p~^3P_{  1  }^o$   & 1.236E+09  &     1.235E+09$^{~(3)}$  &                     &                     \\
$3p^2~^3P_{  1  }$   &   $3s3p~^3P_{  0  }^o$   & 4.148E+08  &     4.144E+08$^{~(3)}$  &                     &                     \\
&   $3s3p~^3P_{  1  }^o$   & 3.058E+08  &     3.062E+08$^{~(3)}$  &                     &                     \\
&   $3s3p~^3P_{  2  }^o$   & 5.170E+08  &     5.167E+08$^{~(3)}$  &                     &                     \\
$3p^2~^3P_{  2  }$   &   $3s3p~^3P_{  1  }^o$   & 3.145E+08  &     3.144E+08$^{~(3)}$  &                     &                     \\
&   $3s3p~^3P_{  2  }^o$   & 9.264E+08  &     9.272E+08$^{~(3)}$  &                     &                     \\
$3p^2~^1S_{  0  }$   &   $3s3p~^1P_{  1  }^o$   & 1.020E+09  &     6.738E+08$^{~(3)}$  &                     &                     \\
&   $3s3p~^3P_{  1  }^o$   & 5.021E+08  &     3.399E+07$^{~(3)}$  &                     &                     \\
$3s3d~^3D_{  2  }$   &   $3s3p~^3P_{  1  }^o$   & 8.977E+08  &     9.072E+08$^{~(3)}$  &                     &                     \\
&   $3s3p~^3P_{  2  }^o$   & 3.019E+08  &     3.046E+08$^{~(3)}$  &                     &                     \\
$3s3d~^3D_{  3  }$   &   $3s3p~^3P_{  2  }^o$   & 1.197E+09  &     1.208E+09$^{~(3)}$  &                     &                     \\
$3s3d~^1D_{  2  }$   &   $3s3p~^1P_{  1  }^o$   & 1.388E+09  &     1.429E+09$^{~(3)}$  & 1.388E+09$^{~(11)}$ &                     \\
$3s4p~^3P_{  0  }^o$ &   $3s4s~^3S_{  1  }$     & 5.639E+07  &     5.705E+07$^{~(3)}$  &                     &                     \\
&   $3s3d~^3D_{  1  }$     & 1.556E+07  &     1.520E+07$^{~(3)}$  &                     &                     \\
$3s4p~^3P_{  1  }^o$ &   $3s4s~^3S_{  1  }$     & 5.649E+07  &     5.724E+07$^{~(3)}$  &                     &                     \\
&   $3s3d~^3D_{  1  }$     & 3.905E+06  &     3.816E+06$^{~(3)}$  &                     &                     \\
&   $3s3d~^3D_{  2  }$     & 1.172E+07  &     1.146E+07$^{~(3)}$  &                     &                     \\
$3s4p~^3P_{  2  }^o$ &   $3s4s~^3S_{  1  }$     & 5.683E+07  &     5.762E+07$^{~(3)}$  &                     &                     \\                     
&   $3s3d~^3D_{  1  }$     & 1.568E+05  &     1.541E+05$^{~(3)}$  &                     &                     \\
&   $3s3d~^3D_{  2  }$     & 2.361E+06  &     2.312E+06$^{~(3)}$  &                     &                     \\
&   $3s3d~^3D_{  3  }$     & 1.319E+07  &     1.294E+07$^{~(3)}$  &                     &                     \\
$3s4p~^1P_{  1  }^o$ &   $3s^2~^1S_{  0  }$     & 1.527E+06  &     0.981E+06$^{~(3)}$  & 5.079E+06$^{~(11)}$ &                     \\
&   $3p^2~^1D_{  2  }$     & 5.835E+07  &     5.897E+07$^{~(3)}$  & 6.307E+07$^{~(11)}$ &                     \\
&   $3s4s~^1S_{  0  }$     & 3.109E+07  &     2.965E+07$^{~(3)}$  & 3.111E+07$^{~(11)}$ &                     \\
\\
$3p3d~^3F_{  2  }^o$ &   $3s3d~^3D_{  1  }$     &  2.956E+08 &     2.07E+08$^{~(4b)}$  & 2.14E+08$^{~(10)}$  &                     \\
$3p3d~^3F_{  3  }^o$ &   $3s3d~^3D_{  2  }$     &  3.174E+08 &     2.19E+08$^{~(4b)}$  & 2.25E+08$^{~(10)}$  &                     \\
$3p3d~^3F_{  4  }^o$ &   $3s3d~^3D_{  3  }$     &  3.794E+08 &     2.47E+08$^{~(4b)}$  & 2.54E+08$^{~(10)}$  &                     \\
$3s4f~^3F_{  2  }^o$ &   $3s3d~^3D_{  1  }$     &  1.981E+08 &     1.97E+08$^{~(4b)}$  & 1.98E+08$^{~(10)}$  &                     \\
$3s4f~^3F_{  3  }^o$ &   $3s3d~^3D_{  2  }$     &  2.096E+08 &     2.09E+08$^{~(4b)}$  & 2.07E+08$^{~(10)}$  &                     \\
$3s4f~^3F_{  4  }^o$ &   $3s3d~^3D_{  3  }$     &  2.360E+08 &     2.35E+08$^{~(4b)}$  & 2.33E+08$^{~(10)}$  &                     \\
$3s5f~^3F_{  2  }^o$ &   $3s3d~^3D_{  1  }$     &  2.801E+07 &     2.40E+07$^{~(4b)}$  & 2.50E+07$^{~(10)}$  &                     \\
$3s5f~^3F_{  4  }^o$ &   $3s3d~^3D_{  3  }$     &  3.438E+07 &     2.85E+07$^{~(4b)}$  & 2.90E+07$^{~(10)}$  &                     \\
$3s6f~^3F_{  2  }^o$ &   $3s3d~^3D_{  1  }$     &  1.957E+07 &     2.90E+07$^{~(4b)}$  & 3.10E+07$^{~(10)}$  &                     \\
&   $3s4d~^3D_{  1  }$     &  1.116E+07 &     1.07E+07$^{~(4b)}$  & 1.00E+07$^{~(10)}$  &                     \\
$3s6f~^3F_{  3  }^o$ &   $3s3d~^3D_{  2  }$     &  1.910E+07 &     3.07E+07$^{~(4b)}$  & 3.30E+07$^{~(10)}$  &                     \\
&   $3s4d~^3D_{  2  }$     &  1.200E+07 &     1.14E+07$^{~(4b)}$  & 1.10E+07$^{~(10)}$  &                     \\
$3s6f~^3F_{  4  }^o$ &   $3s3d~^3D_{  3  }$     &  1.920E+07 &     3.46E+07$^{~(4b)}$  & 3.70E+07$^{~(10)}$  &                     \\
&   $3s4d~^3D_{  3  }$     &  1.367E+07 &     1.28E+07$^{~(4b)}$  & 1.20E+07$^{~(10)}$  &                     \\
\midrule	
\end{tabular}
\tablebib{ $^{(2)}$Present calculations; $^{(3)}$\citet{Froese}; $^{(4b)}$\citet{Butler}; $^{(8)}$\citet{Vujnovic}; $^{(10)}$\citet{ChangWang}; $^{(11)}$\citet{ChangFang}; $^{(12)}$\citet{Kernahan, Smith, Berry, Head}; $^{(13)}$\citet{Trabert, Johnson}.}
\end{table*}		

For all computed E1 transitions in \ion{Al}{II}, the transition data can be found in Table 11. In Table 4, a statistical analysis of the uncertainties to the computed transition rates $A_\text{RCI}$ is performed in a similar way as done for \ion{Al}{I}. The transitions are also arranged here in four groups. Following the conclusions by \citet{Asli} and \citet{thesis}, the transitions involving any of the $3s7p~^{1,3}P$ states have been excluded from this analysis. The discrepancies between the length and velocity forms for transitions including the $3s7p~^{1,3}P$ states are consistently large and thus the computed transition rates are not trustworthy. We note that overall, the average uncertainty, as well as the value that includes $75\%$ of the data appear to be smaller, for each group of transitions, compared to the predicted ones in \ion{Al}{I}. Nevertheless, the maximum values of the uncertainties for the last two groups are larger in comparison to \ion{Al}{I}. This is due to some transitions involving $3p3d~^3F$ as upper level. The strong mixing between the $3p3d~^3F$ and the $3s6f~^3F$ levels results in strong cancellation effects. Such effects often hamper the accuracy of the computed transition data and result in large discrepancies between the length and velocity forms.  

In Table 7, current RCI theoretical transition rates are compared with the values from the MCHF-BP calculations by \citet{Froese} and, whenever available, results from the B-spline configuration interaction (BSCI) calculations by \citet{ChangFang}. For the majority of the transitions, there is an excellent agreement between the RCI and MCHF-BP values with the relative difference being less than $1\%$. Some of the largest discrepancies are observed for the $\{3s3d,3p^2\}~^1D~\rightarrow~3s3p~^{\{1,3\}}P^o$ transitions. According to \citet{Froese}, correlation is extremely important for transitions from such $^1D$ states. In the MCHF-BP calculations, all three types of correlation, i.e. VV, CV and CC, have been accounted for, however the CSF expansions obtained from SD-substitutions are not as large as in the present calculations and the $LS$-composition of the configurations might not be predicted as accurately. Hence, the evaluation of line strengths for transitions involving $^1D$ states and in turn the computation of transition rates involving these states will be affected. Computed transition rates using the BSCI approach are provided for transitions that involve only singlet states. The BSCI calculations do not account for the relativistic interaction and no separate values are given for the different fine-structure components of triplet states. For the $3p^2~^1D~\rightarrow~3s3p~^1P^o$ transition, the discrepancy between the RCI and BSCI values is also quite large. On the other hand, for the $3s3d~^1D~\rightarrow~3s3p~^1P^o$ transition, the BSCI result is in perfect agreement with the present $A_\text{RCI}$ value. The agreement between the current RCI and BSCI transition rates exhibits a broad variation. The advantage of the BSCI approach is that it takes into account the effect of the positive-energy continuum orbitals in an explicit manner. Nevertheless, the parametrized model potential that is being used in the work by \citet{ChangFang} is not sufficient to describe states that are strongly mixed. At last, we note the  discrepancy for the $3s4p~^1P^o~\rightarrow~3s^2~^1S_0$ transition, which is quite large between the RCI and MCHF-BP values and inexplicably large between the RCI and BSCI ones. 

In the singly ionized aluminium, measurements of transition properties are available only for a few transitions. In Table 7, the available experimental results are compared with the theoretical results from the current RCI calculations, as well as the former calculations by \citet{Froese} and \citet{ChangFang}. Transition rates have been experimentally observed for the $3s3p~^{1,3}P^o_1~\rightarrow~3s^2~^1S_0$ transitions in the works by \citet{Kernahan, Smith, Berry, Head} and \citet{Trabert, Johnson}, respectively. In Table 7, the average value of these works is displayed. The agreement with the current RCI results is fairly good. Nonetheless, the averaged $A_\text{obs}$ by \citet{Trabert} and \citet{Johnson} is in better agreement with the value by \citet{Froese}. Additionally, \citet{Vujnovic} provided experimental transition rates for the $3p^2~^1D_2~\rightarrow~3s3p~^1P_1$ and $3p^2~^1D_2~\rightarrow~3s3p~^3P_{1,2}$ transitions by measuring relative intensities of spectral lines. These experimental results, however, differ from the theoretical values. 

In the last portion of Table 7, current transition rates for transitions between states with higher energies are compared with the results from the close coupling (CC) calculations by \citet{Butler} and the early results from the configuration interaction (CI) calculations by \citet{ChangWang}. The results from the latter two calculations are found to be in very good agreement. Furthermore, the agreement between the RCI results and the ones from the CC and CI calculations is also very good for the $3s4f~^3F~\rightarrow~3s3d~^3D$ transitions and fairly good for the $3s5f~^3F~\rightarrow~3s3d~^3D$ transitions. On the other hand, for the $\{3p3d,3s6f\}~^3F~\rightarrow~3s3d~^3D$ transitions, the observed discrepancy between the current RCI values and the ones from the two previous calculations is substantial. This outcome indicates that the calculations by \citet{Butler} and \citet{ChangWang} are insufficient to properly account for correlation and further emphasizes the quality of the present work.

In the same way as for \ion{Al}{I}, the lifetimes of \ion{Al}{II} excited states were also estimated based on the computed E1 transitions. In Table 9, both length $\tau_l$ and velocity $\tau_\upsilon$ forms of the currently computed lifetimes are displayed. As already mentioned, the agreement between these two forms serves as an indication of the quality of the results. The average relative difference between the two forms is $\sim2\%$. The largest discrepancy is observed between the length and velocity gauges of the singlet $3p^2~^1D$ state, as well as the singlet and triplet $3s7p~^{1,3}P$ states. The highest computed levels in the calculations of \ion{Al}{II} correspond to configuration states with orbitals up to $n=7$, such as $3s7p$. Similarly to the conclusions for the lifetimes of \ion{Al}{I}, better agreement between the length and velocity forms of the $3s7p~^{1,3}P$ states could probably be obtained by including in the MR $3snl$ configurations with $n>7$.

In Table 9, the lifetimes from the current RCI calculations are compared with results from previous MCHF-BP and BSCI calculations by \citet{Froese} and \citet{ChangFang}, respectively. Besides the lifetimes of the triplet $3s3p~^3P^o_1$ and singlet $3p^2~^1D_2$ states, the agreement between the RCI and MCHF-BP calculations is very good. Furthermore, the overall agreement between the RCI and BSCI calculations is sufficiently good. Despite the poor agreement between the RCI and BSCI values for the  $3p^2~^1D_2$ state and $3s7p~^{1,3}P$ states, for the rest of the states the discrepancies are of the order of a few per cent. The BSCI results are more extended. However, no separate values are provided for the different $LSJ$-components of the triplet states and for those, the average lifetime is presented instead. 

In Table 9, the theoretical lifetimes are also compared with available measurements. The measured lifetime of the $3s3p~^3P^o_1$ state by \citet{Trabert} and \citet{Johnson} agrees remarkably well with the calculated value by \citet{Froese}. The agreement with the current results is fairly good too. The lifetime of the $3s3p~^1P^o_1$ state measured by \citet{Kernahan}, \citet{Head}, \citet{Berry} and \citet{Smith} is well represented by all theoretical values. On the other hand, the results from the measurements of the $3snf~^3F$ states by \citet{Andersen} differ substantially from the theoretical RCI values. For the $3snf~^3F$ Rydberg series, only theoretical lifetimes using the current MCDHF and RCI approach are available. Being aware of the large uncertainties associated with early beam-foil measurements, the discrepancies between theoretical and experimental values are in some way expected. The only exception is the lifetime of the $3s5f~^3F$ state, which is in rather good agreement with the RCI values. In the experiments by \citet{Andersen} the different fine-structure components have not been separated and a single value is provided for all three different $LSJ$-levels.

\section{Summary and conclusions}

In the present work, updated and extended transition data and lifetimes are made available for both \ion{Al}{I} and \ion{Al}{II}. The computations of transition properties in these two systems are challenging mainly due to the strong two-electron interaction between the $3s3d~^1D$ and $3p^2~^1D$ states, which dominates the lowest part of their spectra. Thus, some of the states are strongly mixed and large amount of correlation is needed to accurately predict their $LS$-compositions. We are confident that in this work enough amount of correlation has been included to affirm the reliability of the results. The predicted excitation energies are in excellent agreement with the experimental data provided by the NIST database, which is a good indicator of the quality of the produced transition data and lifetimes.

We have performed an extensive comparison of the computed transition rates and lifetimes with the most recent theoretical and experimental results. There is a significant improvement in accuracy, in particular for the more complex system of neutral \ion{Al}{I}. The computed lifetimes of \ion{Al}{I} are in very good agreement with the measured lifetimes in high-precision laser spectroscopy experiments. The same holds for the measured lifetimes of \ion{Al}{II} in ion storage rings. The present calculations are extended to higher energies and many of the computed transitions fall in the infrared spectral region. The new generation of telescopes are designed for this region and such transition data are of high importance. The objective of this work is to make available atomic data that could be used to improve the interpretation of abundances in stars. Lists of trustworthy elemental abundances will enable tracing stellar evolution, as well as the formation and chemical evolution of the Milky Way.

The agreement between the length and velocity gauges of the transition operator serves as a criterion to the quality of the transition data, as well as lifetimes. For most of the strong transitions in both \ion{Al}{I} and \ion{Al}{II}, the agreement between the two gauges is very good. For transitions involving states with the highest $n$ quantum number for the $s$ and $p$ symmetries, we observe that the agreement between the length and velocity forms is less good. This becomes more evident when estimating lifetimes of excited levels that are associated with those transitions.

%----------------------Acknowledements------------------
\begin{acknowledgements}
The authors have been supported by the Swedish Research Council (VR) under contract 2015-04842. The authors acknowledge H. Hartman, Malmö University and Lund University, and H. Jönsson, Lund University, for discussions.
\end{acknowledgements}

%-------------------------------------------------------------------

%\bibliography{bibli} 
\bibliographystyle{aa}

\clearpage

\begin{onecolumn}
	
\begin{longtable}{clcrrrrrrrrr}
\caption{\label{energies_alII} Computed excitation energies in cm$^{-1}$ for the 78 lowest states in \ion{Al}{II}. The energies are given as a function of the increasing active set of orbitals, accounting for VV correlation, where $n$ indicates the maximum principle quantum number of the orbitals included in the active set. In the third last column, the final energy values are displayed after accounting for CV correlation. The differences $\Delta E$ between the final computations and the observed values are shown in the last column. The sequence and naming of the configurations and $LSJ$-levels are in accordance with the final (CV) computed energies. The levels of the singlet and triplet $3s6h~^{1,3}H$, as well as the $3p3d~^1D$ level have not yet been observed and thus, no $\Delta E$ values are available for those.} \\
\hline \midrule 
& & & \multicolumn{6}{c}{VV} & & & \\		
Pos. & Conf. & $LSJ$ & $n=8$ & $n=9$ & $n=10$ & $n=11$ & $n=12$ & $n=13$ & CV & $E_\text{obs}$$^{(1)}$ & $\Delta E$ \\
\midrule
\endfirsthead   
\caption{continued.}\\
\hline \midrule       
& & & \multicolumn{6}{c}{VV} & & & \\		
Pos. & Conf. & $LSJ$ & $n=8$ & $n=9$ & $n=10$ & $n=11$ & $n=12$ & $n=13$ & CV & $E_\text{obs}$$^{(1)}$ & $\Delta E$ \\
\midrule
\endhead
\midrule
\endfoot
	
1   &  $3s^2 ~$    & $^1S_{   0  } $   &         0 &          0 &          0 &          0 &          0   &        0  &      0   &       0 &    0 \\
2   &  $3s\,3p~$~  & $^3P_{   0  }^o $ &    36~227 &     36~280 &     36~298 &     36~318 &     36~332   &   36~335  &  37~445  &  37~393 &  -52 \\
3   &              & $^3P_{   1  }^o $ &    36~286 &     36~339 &     36~357 &     36~377 &     36~391   &   36~394  &  37~503  &  37~454 &  -49 \\
4   &              & $^3P_{   2  }^o $ &    36~405 &     36~459 &     36~477 &     36~496 &     36~511   &   36~514  &  37~626  &  37~578 &  -48 \\
5   &              & $^1P_{   1  }^o $ &    59~810 &     59~698 &     59~617 &     59~619 &     59~606   &   59~602  &  59~982  &  59~852 & -130 \\
6   &  $3p^2 ~$    & $^1D_{   2  } $   &    83~542 &     83~596 &     83~620 &     83~641 &     83~657   &   83~660  &  85~692  &  85~481 & -211 \\
7   &  $3s\,4s~$~  & $^3S_{   1  } $   &    89~965 &     90~028 &     90~059 &     90~082 &     90~099   &   90~102  &  91~425  &  91~275 & -150 \\
8   &  $3p^2 ~$    & $^3P_{   0  } $   &    92~679 &     92~709 &     92~716 &     92~736 &     92~750   &   92~752  &  94~211  &  94~085 & -126 \\
\end{longtable}
\tablebib{ $^{(1)}$NIST Atomic Spectra Database 2018 \citep{Kramida}.} \newline \\

\begin{longtable}{clcccccc}
\caption{\label{lifetimes2} Comparison between computed and observed lifetimes $\tau$ in seconds, for 75 excited states in \ion{Al}{II}. For the current RCI calculations both length $\tau_l$ and velocity $\tau_\upsilon$ forms are displayed. In the third and second last columns, the predicted lifetimes from MCHF-BP and BSCI calculations are, respectively, given in length form. The last column contains available lifetimes from experimental measurements, together with their uncertainties.} \\
\hline \midrule 
&  &  & \multicolumn{2}{c}{RCI $^{(2)}$} & MCHF-BP $^{(3)}$ & BSCI $^{(11)}$ & Expt.$^{(12),(13),(14)}$ \\ 
Pos. & Conf. & $LSJ$ & $\tau_l$ & $\tau_\upsilon$ & \multicolumn{2}{c}{$\tau_l$} & $\tau$ \\ 
\midrule
\endfirsthead   
\caption{continued.}\\
\hline \midrule       
&  &  & \multicolumn{2}{c}{RCI $^{(2)}$} & MCHF-BP $^{(3)}$ & BSCI $^{(11)}$ & Expt.$^{(12),(13),(14)}$ \\ 
Pos. & Conf. & $LSJ$ & $\tau_l$ & $\tau_\upsilon$ & \multicolumn{2}{c}{$\tau_l$} & $\tau_\text{obs} $ \\
\midrule
\endhead
\midrule
\endfoot
1  &  $3s3p~$&  $^3P_{  1  }^o$  &   3.274E-04 &  2.965E-04   & 3.051E-04 &            &  3.02~(2)~E-04 $^{(13)}$ \\
2  &         &  $^1P_{  1  }^o$  &   7.120E-10 &  7.089E-10   & 7.141E-10 &  6.70~E-10 &  6.90(13)E-10 $^{(12)}$  \\
3  &  $3p^2~$&  $^1D_{  2  }$    &   3.351E-06 &  2.630E-06   & 2.270E-06 &  2.51~E-06 &                       \\
4  &  $3s4s~$&  $^3S_{  1  }$    &   1.318E-09 &  1.325E-09   & 1.322E-09 &  1.32~E-09 &                       \\
5  &  $3p^2~$&  $^3P_{  0  }$    &   8.091E-10 &  8.032E-10   & 8.098E-10 &            &                       \\
6  &         &  $^3P_{  1  }$    &   8.081E-10 &  8.023E-10   & 8.082E-10 &            &                       \\
7  &         &  $^3P_{  2  }$    &   8.059E-10 &  8.000E-10   & 8.054E-10 &            &                       \\
8  &  $3s4s~$&  $^1S_{  0  }$    &   2.921E-09 &  2.922E-09   & 2.893E-09 &  2.93~E-09 &                       \\
\end{longtable} 
\tablebib{ $^{(2)}$present calculations; $^{(3)}$\citet{Froese}; $^{(11)}$\citet{ChangFang}; $^{(12)}$\citet{Kernahan, Smith, Berry, Head}; $^{(13)}$\citet{Trabert, Johnson}; $^{(14)}$\citet{Andersen}.}

\vspace{1.5cm}

\begin{longtable}{llllll} 
	\caption{Transition data for Al I} \\	
	\hline
	Upper & Lower & 	$\Delta E$ (cm$^{-1}$) & $\lambda$ (\AA) & 	$A$(s$^{-1}$)&$gf$ \\
	\hline
	\\
	\endfirsthead   
	\caption{Continued.}\\
	\hline\hline       
	Upper & Lower & 	$\Delta E$ (cm$^{-1}$) & $\lambda$ (\AA) & 	$A$(s$^{-1}$)&$gf$ \\ 
	\hline
	\\
	\endhead
	\hline
	\endfoot
	
	$3s^2 \,5d~^2D_{ 3/2  }$            	&	 $3s^2 \,3p~^2P_{ 1/2  }^o$          	&	44125	&	2266.251	&	8.20E+07	&	2.53E-01	\\
	$3s^2 \,5d~^2D_{ 5/2  }$            	&	 $3s^2 \,3p~^2P_{ 3/2  }^o$          	&	44025	&	2271.420	&	9.71E+07	&	4.50E-01	\\
	$3s^2 \,5d~^2D_{ 3/2  }$            	&	 $3s^2 \,3p~^2P_{ 3/2  }^o$          	&	44021	&	2271.613	&	1.60E+07	&	4.94E-02	\\
	$3s^2 \,4d~^2D_{ 3/2~b}$            	&	 $3s^2 \,3p~^2P_{ 1/2  }^o$          	&	42232	&	2367.863	&	7.13E+07	&	2.40E-01	\\
	$3s^2 \,4d~^2D_{ 5/2~b}$            	&	 $3s^2 \,3p~^2P_{ 3/2  }^o$          	&	42132	&	2373.437	&	8.41E+07	&	4.26E-01	\\
	$3s^2 \,4d~^2D_{ 3/2~b}$            	&	 $3s^2 \,3p~^2P_{ 3/2  }^o$          	&	42128	&	2373.717	&	1.39E+07	&	4.69E-02	\\
	$3s^2 \,6s~^2S_{ 1/2  }$            	&	 $3s^2 \,3p~^2P_{ 1/2  }^o$          	&	41963	&	2383.014	&	4.83E+06	&	8.22E-03	\\
	$3s^2 \,6s~^2S_{ 1/2  }$            	&	 $3s^2 \,3p~^2P_{ 3/2  }^o$          	&	41859	&	2388.943	&	9.58E+06	&	1.64E-02	\\
\hline
\end{longtable}

\vspace{1.5cm}

\begin{longtable}{llllll} 
		\caption{Transition data for Al II} \\	
		\hline
		Upper & Lower & 	$\Delta E$ (cm$^{-1}$) & $\lambda$ (\AA) & 	$A$(s$^{-1}$)&$gf$ \\
		\hline
		\\
		\endfirsthead   
		\caption{Continued.}\\
		\hline\hline       
		Upper & Lower & 	$\Delta E$ (cm$^{-1}$) & $\lambda$ (\AA) & 	$A$(s$^{-1}$)&$gf$ \\ 
		\hline
		\\
		\endhead
		\hline
		\endfoot
		
		$3s\,7p~^1P_{  1  }^o$ 	&	 $3s^2 ~^1S_{  0  }$        	&	140148	&	713.529	        &	8.07E+06	&	1.85E-03	\\	
		$3s\,6p~^3P_{  1  }^o$ 	&	 $3s^2 ~^1S_{  0  }$        	&	135186	&	739.718	        &	3.30E+04	&	8.12E-06	\\	
		$3s\,6p~^1P_{  1  }^o$ 	&	 $3s^2 ~^1S_{  0  }$        	&	135132	&	740.013	        &	8.37E+06	&	2.06E-03	\\	
		$3s\,5p~^1P_{  1  }^o$ 	&	 $3s^2 ~^1S_{  0  }$        	&	126078	&	793.158	        &	6.16E+06	&	1.74E-03	\\	
		$3s\,5p~^3P_{  1  }^o$ 	&	 $3s^2 ~^1S_{  0  }$        	&	125874	&	794.443	        &	2.44E+04	&	6.92E-06	\\	
		$3s\,4p~^1P_{  1  }^o$ 	&	 $3s^2 ~^1S_{  0  }$        	&	107132	&	933.420	        &	1.53E+06	&	5.99E-04	\\	
		$3s\,6d~^3D_{  1  }$   	&	 $3s\,3p~^3P_{  0  }^o$ 	&	101560	&	984.639	        &	8.34E+06	&	3.64E-03	\\	
		$3s\,6d~^3D_{  1  }$   	&	 $3s\,3p~^3P_{  1  }^o$ 	&	101501	&	985.205	        &	6.19E+06	&	2.70E-03	\\
\hline
	\end{longtable}

\end{onecolumn}

\end{document}